
\input harvmac.tex

\def\inbar{\,\vrule height1.5ex width.4pt depth0pt}
\def\IB{\relax{\rm I\kern-.18em B}}
\def\IC{\relax\hbox{$\inbar\kern-.3em{\rm C}$}}
\def\ID{\relax{\rm I\kern-.18em D}}
\def\IE{\relax{\rm I\kern-.18em E}}
\def\IF{\relax{\rm I\kern-.18em F}}
\def\IG{\relax\hbox{$\inbar\kern-.3em{\rm G}$}}
\def\IH{\relax{\rm I\kern-.18em H}}
\def\II{\relax{\rm I\kern-.18em I}}
\def\IK{\relax{\rm I\kern-.18em K}}
\def\IL{\relax{\rm I\kern-.18em L}}
\def\IM{\relax{\rm I\kern-.18em M}}
\def\IN{\relax{\rm I\kern-.18em N}}
\def\IO{\relax\hbox{$\inbar\kern-.3em{\rm O}$}}
\def\IP{\relax{\rm I\kern-.18em P}}
\def\IQ{\relax\hbox{$\inbar\kern-.3em{\rm Q}$}}
\def\IR{\relax{\rm I\kern-.18em R}}
\font\cmss=cmss10 \font\cmsss=cmss10 at 7pt
\def\IZ{\relax\ifmmode\mathchoice
{\hbox{\cmss Z\kern-.4em Z}}{\hbox{\cmss Z\kern-.4em Z}}
{\lower.9pt\hbox{\cmsss Z\kern-.4em Z}}
{\lower1.2pt\hbox{\cmsss Z\kern-.4em Z}}\else{\cmss Z\kern-.4em Z}\fi}
\def\IGa{\relax\hbox{${\rm I}\kern-.18em\Gamma$}}
\def\IPi{\relax\hbox{${\rm I}\kern-.18em\Pi$}}
\def\ITh{\relax\hbox{$\inbar\kern-.3em\Theta$}}
\def\IOm{\relax\hbox{$\inbar\kern-3.00pt\Omega$}}
\def\tb{\bar{t}}
\def\psib{\bar{\psi}}

\def\CF {{\cal F}}

\def\CO {{\cal O}}

\def\p {\partial}

\def\pb{\bar{\partial}}
\def\FF{{\widetilde{F}}}

\Title{\vbox{\baselineskip12pt\hbox{IASSNS-HEP-92/48}\hbox{YCTP-P22-92}
\hbox{hepth@xxx.9208031}}}
{\vbox{\centerline{The Partition Function of}
\centerline{2D String Theory}}}

\centerline{Robbert Dijkgraaf\foot{Permanent address after August 1:
Department of Mathematics, University of Amsterdam,
Plantage Muidergracht 24, NL-1018 TV Amsterdam, The Netherlands.}}
\centerline{\it School of Natural Sciences}
\centerline{\it Institute for Advanced Study}
\centerline{\it Princeton, NJ 08540}
\medskip

\centerline{Gregory Moore and Ronen Plesser}
\centerline{\it Department of Physics}
\centerline{\it Yale University}
\centerline{\it New Haven, CT 06511-8167}

\bigskip
\noindent
We derive a compact and explicit expression for the generating functional of
all correlation functions of tachyon operators in 2D string theory. This
expression makes manifest relations of the $c=1$ system to KP flow and
$W_{1+\infty}$ constraints. Moreover we derive a Kontsevich-Penner integral
representation of this generating functional.

\Date{August 11, 1992}

\newsec{Introduction}

One of the beautiful aspects of the matrix-model formulation
of $c<1$ string theory is that it gives a natural and
mathematically precise formulation of the partition function
of strings moving in different backgrounds. This result
began with Kazakov's fundamental discovery of the appearance
of matter fields in the one-matrix model
\ref\volodya{V. Kazakov, Mod. Phys. Lett {\bf A4} (1989) 2125.}\
and culminated in the discovery of the generalized KdV flow equations
and the associated $W_N$ constraints in the $c<1$ matrix
models coupled to gravity
\nref\bdss{T. Banks, M.R. Douglas, N. Seiberg, and S. Shenker,
{\it Microscopic and macroscopic loops in nonperturbative two dimensional
gravity}, Phys. Lett. {\bf 238B} (1990) 279.}%
\nref\douglas{M. Douglas, {\it Strings in less than one dimension and the
generalized KdV hierarchies}, Phys. Lett. {\bf 238B} (1990) 176.}%
\nref\grmg{D.J. Gross and A. Migdal, {\it A nonperturbative treatment of two
dimensional quantum gravity}, Nucl. Phys. {\bf B340} (1990) 333.}%
\nref\dvv{R. Dijkgraaf, E. Verlinde, and H. Verlinde, {\it Loop equations and
Virasoro constraints in nonperturbative 2d quantum gravity}, Nucl. Phys.
{\bf B348} (1991) 435.}%
\nref\kawai{M. Fukuma, H. Kawai, and R. Nakayama, {\it Continuum
Schwinger-Dyson equations and universal structures in two-dimensional quantum
gravity}, Int. Jour. Mod. Phys. {\bf A6} (1991) 1385.}%
\refs{\bdss{--}\kawai}.
Recently these results have
been further deepened through the use of a Kontsevich matrix
model representation for the tau functions relevant to
these flows
\ref\kontsevich{M. Kontsevich, {\it Intersection theory on the moduli space of
curves and the matrix Airy function,} Max-Planck-Institut preprint MPI/91-77.},
see also
\nref\morozov{S. Kharchev, A. Marschakov, A. Mironov, A. Morozov, A.
Zabrodin, {\it Unification of all string models with $c\leq 1$}, Phys. Lett.
{\bf 275B} (1992) 311;
{\it Towards unified theory of 2d gravity,}  Lebedev Institute
preprint FIAN/TD-10/91, hepth 9201013\semi
A. Marshakov, {\it On string field theory at $c\le 1$}, Lebedev Institute
preprint FIAN/TD-8/92, hepth 9208022.}%
\nref\robbert{R. Dijkgraaf, {\it Intersection theory, integrable hierarchies
and topological field theory}, IAS preprint IASSNS-HEP-91/91, hepth 9201003.}%
\refs{\morozov,\robbert}.
Analogous results in the $c=1$ model have been strangely absent,
and this paper is a first step in an attempt to change that
situation. Using recently developed techniques
for calculating tachyon correlators in the $c=1$ model
we derive  a simple and compact
expression (equation $(3.10)$)
for the generating functional of tachyon
correlators, or equivalently the
string partition function in an arbitrary tachyon background,
valid to all orders in string perturbation theory. In Euclidean space
this quantity can be interpreted as the partition function
of a nonlinear sigma model as a function of an infinite set of coupling
constants $t_k,\tb_k$ for a set of marginal operators.
Upon appropriate analytic continuation to
Minkowski space the partition function may be interpreted as
the string $S$-matrix in a coherent state basis.

One immediate consequence of our result $(3.10)$ is that
the partition function is naturally represented as a tau function of
the Toda hierarchy. From this result we obtain $W_\infty$
flow equations (equation 4.10) when the $c=1$ coordinate
$X$ is compactified at the self-dual radius.
Moreover, this expression can be
used to derive a Kontsevich-Penner representation
of the partition function as a matrix integral, as
described in section five below. In section six we discuss
how time-independent changes in the matrix model background
fit into our formalism, and in section seven we discuss some
open problems and the relation of this work to other recent
papers on $c=1$ and $W_\infty$.

\newsec{Defining correlation functions}

In a particular background, string propagation in a two-dimensional spacetime
is described on the string worldsheet by the conformal field theory of a
massless scalar $X$ coupled to a $c=25$ Liouville theory $\phi$ with worldsheet
action (excluding ghosts)
\eqn\confld{
\CA=\int \half\p X\pb X + \p\phi\pb\phi +\sqrt{2} R^{(2)}\phi
+\mu e^{\sqrt{2}\phi}
\ .}
Via its dual interpretation as the conformal gauge action for the coupling
of $X$ to two-dimensional gravity, \confld\  is expressible as the continuum
limit of a sum over discretized surfaces.
The discrete sum, as is by now well known, is generated by a matrix integral.
In the double scaling limit which leads to the continuum theory this is in turn
equivalent to a theory of free nonrelativistic fermions with action
\eqn\action{
S=\int_{-\infty}^\infty dx d\lambda\,
 \hat\psi^\dagger\biggl( i{d\over dx}+{d^2\over d\lambda^2}
-V(\lambda)\biggr)\hat\psi\ .
}
The potential $V(\lambda)$ in \action\ is required to approach $-{1\over 4}
\lambda^2$ for large $\lambda$ in order to reproduce the topological
expansion of string theory.%
\nref\mpr{G. Moore, R. Plesser, and S. Ramgoolam, {\it Exact S-matrix for
2D string theory}, Nucl. Phys. {\bf B377}(1992)143; hepth 9111035}%
\foot{The behavior of $V(\lambda)$ for negative $\lambda$ is irrelevant to
all orders of perturbation theory in $1/\mu$. Indeed, the results of this paper
should be interpreted in this perturbative sense. Many results
are true in the nonperturbative context and we will indicate
this in the appropriate places. Where we mention nonperturbative
results we will refer to potentials which grow sufficiently rapidly for large
negative $\lambda$. In \mpr\ these were termed ``type I'' models.}

The theory contains one field theoretic degree of freedom, the massless
``tachyon''. Tachyon correlators are calculated in the theory \confld\ by the
insertion of vertex operators
\eqn\vertnorm{
\CT_q={\Gamma(|q|)\over \Gamma(-|q|)}\int_\Sigma e^{iq X/\sqrt{2}}
e^{\sqrt{2}(1-\half |q|)\phi}
\ .}
In this section we show how these correlators are
calculated in the double scaled matrix model \action.
The presentation is a modification of the
original derivation in
\mpr, which stressed the spacetime interpretation via
collective field theory, in that here we emphasize
the relation to macroscopic loop amplitudes.

We recall that the objects calculated
in a sum over continuous geometries on two-surfaces with
boundary are ``macroscopic loop
amplitudes'' defined by fixing the boundary values of the
two-metric $e^{\sqrt{2}\phi}$ so that the bounding
circles $\CC$ have lengths $\ell=\oint_\CC e^{\phi/\sqrt{2}}$
\nref\ambjorn{J. Ambjorn, J. Jurkiewicz, and Yu. Makeenko, {\it Multiloop
correlators for two-dimen\-sional quantum gravity}, Phys. Lett.
{\bf 251B} (1990) 517.}%
\nref\kostov{I.K. Kostov, {\it Loop amplitudes for non rational string
theories}, Phys. Lett. {\bf 266B} (1991) 317.}%
\nref\mms{E. Martinec, G. Moore, and N. Seiberg,
{\it Boundary operators in 2D gravity}, Phys. Lett. {\bf 263B} 190.}%
\nref\lpstates{G. Moore, N. Seiberg, and M. Staudacher, {\it From loops
to states in two-dimensional quantum gravity},
Nucl. Phys. {\bf B362} (1991) 665.}%
\nref\msi{G. Moore and N. Seiberg, {\it From loops to fields in
2d gravity}, Int. Jour. Mod. Phys. {\bf A7} (1992) 2601.}%
\nref\moore{G. Moore, {\it Double-scaled field theory at $c=1$},
Nucl. Phys. {\bf  B368} (1992) 557.}%
\refs{\bdss,\ambjorn{--}\moore}.
In \refs{\lpstates,\moore} tachyon correlators were defined as the coefficients
of nonanalytic powers of $\ell$ in the small-$\ell$ expansion of macroscopic
loop amplitudes. In particular, in this limit the macroscopic loop
may be written as a sum of local operators
\eqn\locopexp{W_{in}(\ell,p)= -\CT_p {\pi\over sin\pi |p|} \mu^{-|p|/2}
I_{|p|}(2\sqrt{\mu}\ell)-
\sum_{r=1}^\infty\hat\CB_{r,p}{2(-1)^r r\over r^2-p^2}
\mu^{-r/2}I_r(2\sqrt{\mu}\ell)
}
where $\hat\CB_{r,p}$ are redundant operators for $p\notin \IZ$.
We may thus extract tachyon correlators from macroscopic loop amplitudes as
\eqn\tacslps{
\langle \prod_{i=1}^n W(\ell_i,q_i) \rangle =
\prod_{i=1}^n \Gamma(-|q_i|)\ell_i^{|q_i|}
\bigl( \langle \prod_{i=1}^n \CT_{q_i} \rangle + \CO (\ell_i^2) \bigr) +
\hbox{ analytic in }\ \ell_i\ .
}
The matrix model formulation of the theory leads to a simple computation of the
appropriate limits of loop amplitudes.
In the matrix model the macroscopic loop is related by a Laplace transform
to the eigenvalue density
$\hat \rho(\lambda,x) = \hat \psi^{\dagger} \hat \psi (\lambda,x)$:
\eqn\wro{\eqalign{
W(\ell,x) &= \int_0^{\infty} e^{-\ell \lambda} \hat \rho(\lambda,x) d\lambda\cr
W(\ell,q) &= \int_{-\infty}^{\infty} e^{iqx} W(\ell,x) dx \ .\cr
}}
Defining
\eqn\wlplc{
\hat W(z,x) = \int_0^{\infty} e^{-z\ell} W(\ell,x) d\ell
}
we recover
\eqn\roagain{
\hat \rho(\lambda,x) = -{i \over \pi} {\rm Disc}\,(\hat W(z,x))
\big|_{z = -\lambda}\ .
}
Inserting \tacslps\ we find now
\eqn\rotacs{
\langle \prod_{i=1}^n \hat \rho(\lambda_i,q_i) \rangle =
\prod_{i=1}^n \lambda_i^{-q_i-1} \Bigl(
\langle \prod_{i=1}^n \CT_{q_i} \rangle + \CO (\lambda_i^{-2}) \Bigr)
}
for generic $q_i$.
In the next section we show how this leads to a simple formula
for the correlation functions.

\newsec{Calculating correlators}

\subsec{Graphical rules for tachyon correlators}

The calculation of tachyon correlation functions now reduces to
the study of asymptotics of correlation functions of the eigenvalue density.
An explicit formula for these was found in \moore;
we will use a generalization of this
to a compactification radius $\beta$
for the Euclidean $X$ field. Momenta for the tachyon
field are therefore always of the form $q=n/\beta$ for $n\in \IZ$.
The required modification follows by interpreting the compactified Euclidean
time as a finite temperature for the free fermion system
\nref\kleblow{I.R. Klebanov and D. Lowe, {\it Correlation functions in
two-dimensional gravity coupled to a compact scalar field}, Nucl. Phys.
{\bf B363} (1991) 543.}%
\kleblow.
Allowed fermionic momenta are of the form $p_m=(m+\half)/\beta$ with
$m\in\IZ$.

The eigenvalue correlator can be written as:
\eqn\nptev{
\langle \prod \hat\rho(\lambda_i,q_i)\rangle=\sum_{m=-\infty}^\infty
\sum_{\sigma\in\Sigma_n}
\prod_{k=1}^n I(Q^\sigma_k,\lambda_{\sigma(k)},\lambda_{\sigma(k+1)})
}
where $\Sigma_n$ is the set of permutations of $n$ objects,
$Q_k^\sigma\equiv p_m+q_{\sigma(1)}+\cdots + q_{\sigma(k)}$.
The $\lambda$ dependence of the correlator is determined by the function
$I(q,\lambda_1,\lambda_2)=(I(-q,\lambda_1,\lambda_2))^*=\langle\lambda_1
|{1\over H-\mu-iq}|\lambda_2\rangle$ where $H$ is the one-body Hamiltonian
for the free fermion system. This has large $\lambda$ asymptotics given by
a direct and a reflected contribution \mpr
\eqn\asympi{\eqalign{
I(q,\lambda_1,\lambda_2)&
{\buildrel \lambda\to +\infty\over\sim}{-i\over\sqrt{\lambda_1\lambda_2}}
\biggl[D[q,\lambda_1,\lambda_2]+R[q,\lambda_1,\lambda_2]\biggr]\cr
D[q,\lambda_1,\lambda_2]&= exp\biggr[ i(\mu+iq) |\log {(\lambda_1/\lambda_2)}|
-
{i\over 4} |\lambda_1^2 - \lambda_2^2|\biggr]\cr
R[q,\lambda_1,\lambda_2]&= R_q \, exp\biggr[
i(\mu+iq)\log{(\lambda_1\lambda_2)}
-{i\over 4} (\lambda_1^2+\lambda_2^2)\biggr]\ ,\cr
}}
for $q > 0$. The function $R_q$ is the reflection coefficient for the
nonrelativistic free fermions in the double scaled potential $V$. For the
``standard'' case $V = -{1\over 4}\lambda^2$ with an
infinite wall at $\lambda=0$ it is given by
\eqn\rq{
R_q =i\sqrt{1+i e^{-\pi(\mu+iq)}\over
1-i e^{-\pi(\mu+iq)}}
\sqrt{\Gamma(\half-i\mu+q)\over\Gamma(\half+i\mu-q)}
\ .}
Inserting \asympi\ in the expression \nptev\ leads to a sum of terms.
The calculation of tachyon correlators requires the extraction of those
terms in the sum with the correct asymptotic dependence on $\lambda_i$. For
each permutation $\sigma$, at most a finite number of terms in the
sum over the loop momentum $p_m$ contribute to the result.

A graphical procedure for performing this extraction was developed in \mpr\ and
used to derive an explicit expression for arbitrary tachyon correlators.
We divide the tachyon insertions into ``incoming'' ($q<0$) and `outgoing''
($q>0$) particles.
As in a Feynman diagram there is a vertex in the $(x,\lambda)$ half-space
corresponding to each operator $\hat\rho(x,\lambda)$.
While the final result will of course be independent of the order in which
the $\lambda_i$ are increased to infinity, in intermediate steps we will
choose some order and locate the vertices accordingly.
Points are connected by line segments, representing the integral
$I$, to form a one-loop graph. Since the expression for $I$ in
\asympi\  has two terms we have both direct
and reflected propagators as in
\fig\idiagram{(a) A pictorial version of the integral $I$ for
positive momentum. (b) A pictorial version of the integral $I$ for
negative momentum}.
Each line segment carries a momentum and an arrow. Note that
in \idiagram\ the reflected propagator, which we call simply
a ``bounce,'' is composed of two segments with opposite
arrows and momenta. These line segments are joined to form
a one-loop graph according to the following rules:

\medskip

{\parindent=1truein
\item{RH1.} Lines with positive (negative) momenta slope upwards to the
right (left).

\item{RH2.} At any vertex arrows are conserved and momentum
 is conserved as
time flows upwards. In particular momentum
$q_i$ is inserted at the vertex in
\fig\vertex{Incoming and outgoing vertices.
The dotted line carrying negative (positive) momentum $q_i$ should be
thought of as an incoming (outgoing) boson with energy $|q_i|$.
Momentum carried by lines is always conserved as time flows upwards.}.

\item{RH3.} Outgoing vertices at $(x_{out},\lambda_{out})$ all have
later times than incoming vertices $(x_{in},\lambda_{in})$: $x_{out}>x_{in}$.

}

\medskip

Diagrams drawn according to these rules correspond to possible physical
processes in real time and were hence termed ``real histories''.
The connected tachyon correlation function is found by summing the terms in
\nptev\ corresponding to all real histories, and reads schematically
\eqn\radmis{
\langle \prod_{i=1}^n \CT_{q_i} \rangle =
(-i)^n \sum_{\rm RH}\pm \sum_m
\prod_{\rm bounces} R_Q (-R_Q)^*\ .
}
The graphical rules allow one to convert \radmis\ into an explicit formula
for the amplitude \mpr. In the next subsection we will show that this
result may be written quite simply in terms of free fermionic fields,
representing a fermionized version of the free relativistic bosonic field
which describes the asymptotic behavior of the tachyon.

\subsec{Free Energy in terms of free oscillators}

One of the central results of \mpr\ is that the graphical rules described
above are equivalent to the composition of three transformations on the
scattering states: fermionization, free fermion
scattering, and bosonization: $i_{f\to b}\circ S_{ff}\circ i_{b\to f}$
as in
\fig\composition{A real history as a composition of three maps}.
The various real histories correspond to the possible contractions among the
incoming and outgoing fermions, and the fermion scattering matrix describes
a simple one-body process, given essentially by the phase shift in the
nonrelativistic problem.
It should be noted that this does not imply the (false) statement that
bosonization is exact for the nonrelativistic fermion problem. Rather, it
is a statement about the {\it  asymptotics} of certain correlators in the
theory for a particular class of potentials. Here we will rewrite the tachyon
amplitude using this formulation as a matrix element of a certain operator in
the conformal field theory of a free Weyl fermion.

It is convenient to define rescaled tachyon vertex operators
$V_q = \mu^{1-|q|/2} \CT_q$ and two free scalar fields
$\p\phi^{in/out} = \sum_n \alpha_n^{in/out} z^{-n-1}$, such that
\eqn\bosn{
\langle \prod_{i=1}^n V_{n_i/\beta} \prod_{j=1}^m V_{-n'_j/\beta} \rangle =
-{(i\mu)^n \over \beta}\langle 0|\prod \alpha^{out}_{n_i} \prod
\alpha^{in}_{-n_i'}|0\rangle
}
where $|0\rangle$ is the standard $SL(2,\IR)$ invariant vacuum.
The equivalence of the graphical rules to bosonization then implies that
while the relation between the two bosonic fields is complicated and
nonlinear it may in fact be expressed as a simple linear transformation
in the fermionized version.
Thus we write
$\p \phi= \psi(z)\bar\psi(z)$ where $\psi,\bar\psi$ are
Weyl fermions of weight $\half$ with expansions
\eqn\expwyl{
\eqalign{
\psi(z)&=\sum_{m\in\IZ} \psi_{m+\half} z^{-m-1}\cr
\psib(z)&=\sum_{m\in\IZ} \psib_{m+\half} z^{-m-1}\cr
\{\psi_{r},\psib_s\}&=\delta_{r+s,0}\ .\cr
}}
Now the result of \mpr\ states that \bosn\ is equivalent to
\eqn\trans{\eqalign{
\psi^{in}_{-(m+\half)}&=R_{p_m}\psi^{out}_{-(m+\half)}\cr
\bar\psi^{in}_{-(m+\half)}&=R^*_{p_m}\bar\psi^{out}_{-(m+\half)}\ .\cr
}}
Unitarity of the tachyon $S$-matrix is equivalent to the identity
\eqn\unident{
R_q R^*_{-q} = 1}
on the reflection factors.%
\foot{This holds to all orders in perturbation theory for any of the potentials
we consider. The question of its nonperturbative validity was discussed in
\mpr. Essentially, this requires that $V(\lambda)$ grow sufficiently
rapidly for large negative $\lambda$.}
Using this, we can rewrite \trans\ as a unitary
transformation
\eqn\unitran{\eqalign{
\psi^{in}(z)&=S\psi^{out}(z) S^{-1}\cr
\psib^{in}(z)&=S\psib^{out}(z) S^{-1}\cr
S&=:exp\biggl[\sum_{m\in\IZ} \log
R_{p_m}\psi^{out}_{-(m+\half)}\psib^{out}_{m+\half}\biggr]:\ .\cr
}}
Thus we may write the full generating
functional for connected Green's functions in terms of
a {\it single} free boson with modes $\alpha_n$:
\eqn\genfun{\eqalign{
\mu^2 \CF &\equiv
\langle e^{\sum_{n\geq 1} t_n V_{n/\beta} +\sum_{n\geq 1} \bar t_n
V_{-n/\beta}}
\rangle_c\cr
&={-1 \over \beta}\langle 0|e^{i\mu \sum_{n\geq 1} t_n \alpha_{n} } S
e^{i\mu \sum_{n\geq 1} \bar t_n \alpha_{-n} }|0\rangle_c\ .\cr
}}
With this definition $\CF$ has a genus expansion
$\CF=\CF_0 + {1\over \mu^2} \CF_1+\cdots$.
This formula is an enormous simplification over previous
expressions for $c=1$ amplitudes. The generating function
for all amplitudes is
\eqn\disc{
Z=e^{\mu^2\CF}
\ .}

\newsec{$W_{1+\infty}$ constraints}

In correlation functions of tachyons with integer (Euclidean) momentum, the
bounce factors $R_q$ of \rq\ simplify due to the following identity
\eqn\useful{
R_{\xi}^* R_{n-\xi}=(-i\mu)^{-n} \bigl(\half-i\mu-\xi\bigr)_n
\ .}
This is valid to all orders in perturbation theory.
(In the ``standard'' potential it also holds
nonperturbatively if $n \in 2\IZ$.)
Note that equation \rotacs\ holds for generic momenta;
the results for integer momenta are defined  by continuity.
Working with the generating functional of all
amplitudes \disc\ we have:
\eqn\difz{\eqalign{
 {i\beta\over\mu}{\p Z\over \p \tb_n}=&
\langle 0|e^{i\mu \sum_{n\geq 1} t_n \alpha_{n} } (S \alpha_{-n} S^{-1}) S
e^{i\mu \sum_{n\geq 1} \bar t_n \alpha_{-n} }|0\rangle\cr
= \oint {dw\over w^n}\oint {dz\over z} &
\biggl[\sum_{m\in\IZ} R_{p_m} R^*_{n/\beta-p_m} ({w\over z})^m
\biggr]\cr
\langle 0|&e^{i\mu \sum_{n\geq 1} t_n \alpha_{n} } \psi(z)\bar\psi(w)
S  e^{i\mu \sum_{n\geq 1} \bar t_n \alpha_{-n} }|0\rangle\cr}
}
At the self-dual radius $\beta=1$,
where all tachyon momenta are integral, we may simplify the
sum on $m$ using \useful
\eqn\sumonm{
(-i\mu)^{-n}(-i\mu+z{\p\over \p z})_n
\sum_{m\in\IZ}  ({w\over z})^m
}
the latter sum acting like a delta function.
Now integrate by parts and use the identity
\eqn\simdfop{
(-i\mu-z{\p\over \p z})_n =(-z)^n z^{-i\mu} ({\p\over \p z})^n z^{i\mu}
\ .}
It is convenient to bosonize
$\psi(z)=e^{\phi(z)}$,$\bar \psi(z)=e^{-\phi(z)}$
and shift the zero mode:
\eqn\shifmod{
\tilde \phi(z)=\phi(z)+i \mu \log z
\ .}
Taking the operator product of the two exponentials in
$\phi$, and using the delta function and charge conservation we find
the operator:
\eqn\intstge{
\oint dw (i\mu)^{-n} {1\over n+1}: e^{-\tilde \phi(w)} \p_w^{n+1}
e^{\tilde \phi(w)}:
\ .}
Now go to the coherent state basis in the $t_n$'s, and redefine
the scalar field by a factor of $i\mu$ to obtain the final result:
\eqn\wardid{
{\p \CF\over \p \tb_n}=Z^{-1}\oint dw {(i \mu)^{-(n+1)}\over n+1}
:e^{-i\mu \varphi(w)}\p_w^{n+1} e^{i\mu \varphi(w)}: Z
}
where
\eqn\finalphi{
\p \varphi(w)={1\over w} +\sum_{n>0} n t_n w^{n-1} -{1\over \mu^2}
\sum_{n>0} {\p\over\p t_n} w^{-n-1}
\ .}
The genus zero result of
\nref\mrpl{G. Moore and R. Plesser, {\it Classical scattering in 1+1
dimensional string theory}, Yale preprint YCTP-P7-92, hepth 9203060, to
appear in Phys. Rev. D.}%
\mrpl\ is easily obtained from this as the leading term at large $\mu$.
(Note that this was obtained at $\beta=\infty$ but genus zero correlators are
independent of $\beta$ \kleblow.)

The operators $P^{(n)}(z)= :e^{-\tilde\phi(z)}\p^n e^{\tilde\phi(z)}:$ and
their derivatives generate the algebra $W_{1+\infty}$
\ref\kawaii{M. Fukuma, H. Kawai, and R. Nakayama, {\it Infinite dimensional
grassmannian structure of two-dimensional quantum gravity}, Commun. Math.
Phys. {\bf 143} (1992) 371.}. The standard generators are related to these by
\eqn\ptow{
W^{(n)}(z) = \sum_{l=0}^{n-1} {(-1)^l\over (n-l) l!}
{\left( (n-l)_l\right)\over(2n-2)_l} \p^l P^{(n-l)}(z)
\ .}
The rescaling of the scalar field required to obtain \finalphi\ is simply a
change of basis effected by the operator $:e^{\log (i\mu) \pi_\phi
\tilde\phi}:$ where $\pi_\phi$ is the momentum conjugate to $\tilde\phi$.
Inserting this we can rewrite \wardid\ as
\eqn\wconstr{\eqalign{{\p Z\over\p\tb_n}&=W^{(n+1)}_{-n} Z\cr
W^{(n+1)}_{-n} &\equiv \oint dz W^{(n+1)}(z) \ .\cr}
}
Generalizations of \wconstr\ to other radii $\beta\not= 1$
follow from \difz.

In the conclusions we comment briefly on the relation of
this result to the many other occurrences of $W_\infty$ in
this subject.

\newsec{Tau-functions and the Kontsevich-Penner matrix integral}

\nref\penner{R.C. Penner,  Commun. Math. Phys. {\bf 113} (1987) 299;
J. Diff. Geom.{\bf 27} (1988) 35.}%
In this section we will point out that the above reformulation of the
generating functional of the $c=1$ string represents mathematically a
$\tau$-function of the Toda Lattice hierarchy. The Toda Lattice naturally
contains the KP and KdV hierarchies, and thus the $c=1$ results are closely
related to the expressions obtained for $c<1$.  We will also show how to
rewrite the partition function (at the self-dual  radius) as a matrix integral,
generalizing expressions previously considered  by Kontsevich \kontsevich\ and
Penner \penner.

\subsec{Grassmannians and tau-functions}

Let us first briefly explain the notion of a tau-function and its relation
with the universal Grassmannian. For more details see {\it e.g.} \ref\segal{G.
Segal and G. Wilson, {\it Loop groups and equations of KdV type,} Publ. Math.
I.H.E.S. {\bf 61} (1985) 1.}  and \ref\date{E. Date, M. Jimbo, M. Kashiwara, T.
Miwa, {\it Transformation  groups for soliton equations,} RIMS Symp. {\sl
Nonlinear Integrable  Systems---Classical Theory and Quantum Theory} (World
Scientific,  Singapore, 1983).}. We will focus here on the relation with
conformal field theory instead of the Lax pair formulation.

Consider a two-dimensional free chiral scalar field $\varphi(z)$, with the
usual mode expansion
\eqn\modes{
\partial\varphi(z) = \sum_n \alpha_n z^{-n-1}\ .
}
The reader is encouraged to think about this scalar field as the target space
tachyon field at spatial infinity with a periodic Euclidean time coordinate.
We have a Hilbert space ${\cal H}$ built on the vacuum $|0\rangle$, and
as in the case of a harmonic oscillator one can consider coherent
states,
\eqn\coh{
|t\rangle = \exp \sum_{n=1}^\infty it_n \alpha_{-n}
}
and their Hermitian conjugates
\eqn\cnj{
\langle t | = \langle 0| \exp \sum_{n=1}^\infty -it_n \alpha_n
}
(The parameters $t_n$ are considered to be real here.)
Now to any state $|W\rangle$ in the Hilbert space ${\cal H}$ we can
associate a coherent state wavefunction $\tau_W(t)$ by considering the inner
product
\eqn\wave{
\tau_W(t) = \langle t|W\rangle\ .
}
This function is a tau-function of the KP hierarchy if and only if
the state $|W\rangle$ lies in the so-called Grassmannian.

To explain the concept of the Grassmannian we have to turn to the
alternative description of this chiral conformal field theory
in terms of chiral Weyl fermions $\psi(z)$,
$\psib(z)$ by means of the well-known bosonization formulas%
\foot{Since we do not wish to flaunt tradition we change conventions
for bosonization in this section relative to the previous sections.}
\eqn\bos{
i\partial\varphi = \psib\psi,\quad \psi =
e^{i\varphi},\quad \psib =e^{-i\varphi}\ .
}
Loosely speaking, the Grassmannian can be defined as the collection of all
{\it fermionic} Bogoliobov transforms of the vacuum $|0\rangle$.
That is, the state $|W\rangle$ belongs to the Grassmannian if it is
annihilated by particular linear combinations of the fermionic creation
and annihilation operators.
\eqn\ann{
(\psi_{n+{1\over2}} - \sum_{m=1}^\infty
A_{nm} \psi_{-m+{1\over2}} )|W\rangle=0,\qquad n\geq 0\ ,
}
or equivalently,
\eqn\bog{
|W\rangle = S \cdot |0\rangle, \qquad S =
\exp \sum_{n,m} A_{nm} \psib_{-n-{1\over2}}\psi_{-m+{1\over2}}\ .
}
Note that the operator $S$ can be considered as an element of the
infinite-dimensional linear group, $S \in GL(\infty,\IC)$.

By replacing the vacuum $|0\rangle$ by the state $|W\rangle$, we simply made
another decomposition into positive and negative energy states, and filled
these new negative energy states.
The positive energy wave-functions are no longer of the form $z^n$ ($n \geq 0$)
but are now given by the functions
\eqn\sta{
v_n(z) = z^n - \sum_{m=1}^\infty A_{nm}z^{-m}\ .
}
If one prefers the language of semi-infinite differential forms, we have
a formula of the form
\eqn\halfinf{
|W\rangle=v_0 \wedge v_1\wedge \ldots
}
In the above fashion one generates solutions to the KP hierarchy.
This construction can be extended to give a tau-function for the Toda
Lattice hierarchy by considering a second set of times $\tb_k$, as discussed
in detail in \ref\toda{K. Ueno and K. Takasaki, {\it Toda lattice hierarchy,}
in
{\sl Group Representations and Systems of Differential Equations,} H. Morikawa
Ed., Advanced Studies in Pure Mathematics 4.}. In terms of our conformal field
theory, the {\it Toda tau-function} is simply obtained as
\eqn\tl{
\tau(t,\tb) = \langle t | S | \tb \rangle\ ,
}
with $|\tb\rangle$ and $\langle t |$ the coherent states \coh\ and \cnj\ and
$S$ a general $GL(\infty,\IC)$ element, {\it i.e.} an exponentiated fermion
bilinear of type \bog.
The $S$-matrix \unitran\ of the $c=1$ string is definitely of this
form, which allows us to conclude that the string partition function can
indeed (after a rescaling $t_n \rightarrow \mu \cdot t_n$) be identified as a
Toda Lattice tau-function.

Instead of taking the inner product of the state $|W\rangle$ with a coherent
bosonic state, one can also consider {\it fermionic} $N$-point functions
(see {\it e.g.} \robbert).
In fact, one finds in this way a simple expression in terms of an $N \times N$
determinant of the wave-functions \sta\
\eqn\fercor{
\langle N | \psi(z_1)\ldots\psi(z_N)|W\rangle = \det v_{j-1}(z_i)\ .
}
Using the bosonization formulas, one recognizes this correlation function
as a special coherent state where the parameters $t_n$ are given by
\eqn\tns{
t_n = \sum_{i=1}^N {1\over n} z_i^{-n}\ .
}
With this choice of parameterization, and after taking into account a normal
ordering contribution, the tau-function can be written as
\eqn\tfn{
\tau(t) = {\det v_{j-1}(z_i)\over \Delta(z)}\ ,
}
with $\Delta(z)$ the Vandermonde determinant $\Delta(z) = \det z_i^{j-1} =
\prod_{i>j} (z_i-z_j).$ We apply this result in the next section.

\subsec{Kontsevich integrals and the $c<1$ models.}

Since our expression for the $c=1$ partition function is very analogous to the
result found for the $c<1$ string theories, we will briefly summarize
the latter (see \refs{\bdss,\douglas,\grmg}.)
Recall that the $c<1$ matrix models naturally give rise to
a universal set of observables $\CO_n$  ($n=1,2,\ldots$) whose
correlation functions
\eqn\cor{
\langle  \CO_{n_1}\cdots \CO_{n_s}\rangle_g
}
at a specific genus $g$ are unambiguously determined.
The generating functional $\tau(t)$ of these
correlators has an asymptotic expansion in the string coupling constant
$\lambda$
\eqn\gen{
\log \tau(t) = \sum_{g=0}^\infty \lambda^{2-2g} \langle \exp \sum_{n=1}^\infty
t_n \CO_n \rangle_g\ .
}
The techniques of the double-scaled matrix models leads to two important
results. First, the partition function $\tau(t)$ is a tau-function of the KP
hierarchy, that is,  it can be written as
\eqn\tauw{
\tau(t) = \langle t | W \rangle =  \langle t | S | 0 \rangle\ ,
}
for some state $|W\rangle$ and matrix $S \in GL(\infty,\IC)$.
Secondly, all minimal models of type $(p,q)$ with fixed $p$ belong
to one  KP orbit. More precisely, relative to a convenient  choice of origin,
the $(p,q)$ model is obtained at the value $t_k=\delta_{k,p+q}$. Furthermore,
the KP hierarchy reduces to the $p^{th}$ KdV hierarchy, which implies that all
correlation functions of the operators $\CO_n$ with $n \equiv 0$ (mod $p$)
vanish.

The state $|W\rangle$ corresponding to this orbit is most simply
described at the $(p,1)$ point, where a description in terms of topological
field theory can be given. For its basis one can take the wave-functions
\eqn\basis{
v_n(z) = \sqrt{ipz^{p-1}\over 2\pi \lambda}  e^{{ip\over p+1}z^{p+1}/\lambda}
\cdot \int_{-\infty}^{\infty} \!dy
\cdot y^n\cdot e^{i(z^py-{y^{p+1}\over p+1})/\lambda}\ ,
}
where the normalization is chosen such that we have the appropriate
asymptotic expansion
\eqn\asy{
v_n(z) = z^n(1 + O(z^{-1}))\ .
}
Since the wave-functions are moments in a Fourier transform, the
fermionic formula \tfn\ can be explicitly evaluated, and gives rise to a
so-called Kontsevich integral \kontsevich\ (see also
\nref\izii{C. Itzykson and J.-B. Zuber, {\it Combinatorics of the
modular group II: The Kontsevich integrals}, Saclay preprint
SPHT/92-001, hepth 9201001\semi
Ph. Di Francesco, C. Itzykson and J.-B. Zuber,
{\it Polynomial averages in the Kontsevich model}, hepth 9206090}%
\refs{\morozov,\robbert,\izii})
\eqn\matrix{
\tau(t) = c \cdot  \int DY \cdot e^{iTr(Z^pY - {Y^{p+1}\over
p+1})/\lambda}\ .
}
Here $Y$ and $Z$ are both $N \times N$ Hermitian matrices, and
the parameterization of the KP times $t_k$ in terms of the matrix $Z$ is
\eqn\para{
t_k = {\lambda\over k} Tr Z^{-k}\ .
}
This result can be generalized to the `generalized Kontsevich model' \morozov\
which features an arbitrary potential $V(z)$
\eqn\gkm{
\tau(t) = c(Z) \cdot \int DY \cdot e^{iTr(V'(Z)Y - V(Y))/\lambda}\ .
}
with
\eqn\norm{
c(Z) = (2\pi i/\lambda)^{-{N^2\over 2}}
\cdot \det V''(Z) \cdot {\Delta(V'(z)) \over \Delta(z)} \cdot
e^{iTr(V(Z)-V'(Z) Z)/\lambda}\ .
}
It has been noticed by many authors that the case $p=-1$ ({\it i.e.} a
logarithmic potential $V(z) = \log z$) is likely associated with the $c=1$
model
\ref\logkon{See, e.g., E. Witten, {\it The N matrix model
and gauged WZW models}, IAS preprint IASSNS-HEP-91/26}.
We will now proceed to show that this is indeed the case.

\subsec{The Kontsevich-Penner integral}

We have seen that the $c=1$ partition function can be succinctly written as a
tau-function of the Toda Lattice hierarchy
\eqn\tfnct{
\tau(t,\tb) = \langle t | S | \tb \rangle\ .
}
 For {\it fixed} $\tb_k$ we recover a tau-function of the KP hierarchy,
which we can study with the techniques of the previous subsection.
Indeed the operators $\CO_n$ of the minimal models should now be compared to
the outgoing tachyons of the $c=1$ model.

We want to determine in more detail the element $W(\tb)$ in the Grassmannian
that parametrizes this particular orbit of the KP flows.
To this end we have to consider the state
\eqn\state{
|W(\tb)\rangle = S \cdot U(\tb) \cdot |0\rangle,\qquad
U(\tb)= \exp \sum_{n=1}^\infty i\mu \tb_n \alpha_{-n}\ .
}
We will describe $|W(\tb)\rangle$ by giving a basis
$v_k(z;\tb)$, $k \geq 0$, of one-particle wave-functions. First we observe that
the operator $U(\tb)$ acts on the wave-functions $z^n$ by simple multiplication
\eqn\uaction{
U(\tb) : \ z^n \rightarrow \exp\left(\sum i\mu \tb_k z^{-k}\right)\cdot z^n\ .
}
Similarly we have for the action of $S$ a multiplication
\eqn\saction{
S: \ z^n \rightarrow R_{p_n} \cdot z^n\ .
}
We have already seen that the reflection factors $R_{p_n}$ contain all the
relevant information of the $c=1$ matrix model. At radius $\beta$ they
can be chosen to be
\eqn\refl{
R_{p_n} = (-i\mu)^{-{n+{1\over2}\over \beta}}
{\Gamma(\half - i\mu + {n+{1\over2}\over \beta}) \over \Gamma(\half -i\mu)}\ .
}
(Recall, we are only interested in the perturbative part in $\mu^{-1}$
of this expression.)
The usual vacuum $|0\rangle$ is spanned by the non-negative
powers $z^k$. Therefore the basis elements $v_k(z;\tb)$ of $W$ are simply
determined as
\eqn\basis{
v_k(z;\tb) = c_k \cdot S \circ U(\tb) z^k\ ,
}
with a normalization constant $c_k$ such that $v_k(z;0) = z^k$. (This
corresponds to the normal ordering of the $S$-matrix in \unitran.)
Since the reflection factor is basically a gamma function, the result can be
expressed as a Laplace transform
\eqn\integral{
v_k(z;\tb) = c'(z) \cdot \int_0^\infty dy \cdot y^k \cdot
y^{-i\mu\beta +(\beta-1)/2} e^{i\mu (y/z)^\beta}
\exp \left(\sum i\mu\tb_k y^{-k}\right)
}
Here the constant $c'(z)$ is given by
\eqn\cnst{
c'(z) = \beta {(-i\mu/z^\beta)^{{1\over2}-i\mu} \over \sqrt{z}\Gamma
(\half - i\mu)}\ .
}
These integral representations are of Kontsevich type if and only if
$\beta=1$, that  is, only at the self-dual radius.
Indeed in that case we have
\eqn\betaone{
v_k(z;\tb) = c'(z) \cdot \int_0^\infty dy \cdot y^k \cdot
\exp i\mu \left(y/z - \log y +\sum \tb_k y^{-k}\right)}
Therefore, following the procedure in \refs{\morozov,\robbert},
we can write the following matrix integral representation for
the generating functional. Define
the integral
\eqn\sig{
\sigma(Z,\tb)= \int dY e^{i\mu Tr[Y Z^{-1} + V(Y)]}\ ,
}
where
\eqn\pot{
V(Y) =-\log Y + \sum \tb_k Y^{-k}\ ,
}
and we integrate over positive definite matrices $Y$. Then we have
\eqn\bloep{
\tau(t,\tb)={\sigma(Z,\tb) \over \sigma (Z,0)}\ ,
}
with the parameterization
\eqn\floep{
t_n = \mu^{-1} \cdot {1\over n} Tr Z^{-n}\ .
}
Note that with this normalization $\tau(t,0)=1$, which is appropriate since we
consider normalized correlation functions.  In order to write down the result
\bloep\  we had to treat the incoming and outgoing tachyons very differently,
parametrizing the outgoing states through \floep, whereas the coupling
coefficients to the incoming states enter the matrix integral in a much more
straightforward fashion. Equation \bloep\ should be considered
as an asymptotic expansion in $\mu^{-1}$, but, for small enough
$t_k,\tb_k$ the expansion in these variables will be
convergent. In some cases, (e.g. the sine-Gordon case considered in
\nref\sg{G. Moore, {\it Gravitational phase transitions and the
sine-Gordon model}, Yale preprint YCTP-P1-1992, hepth 9203061.}%
\sg\ )
the expansion has a finite radius of convergence, and as we increase
$|t_k|$ beyond the radius of convergence we can have phase transitions.

\subsec{The partition function}

\nref\harer{J. Harer and D. Zagier, Invent. Math. {\bf 185}  (1986) 457.}%
\nref\iz{C. Itzykson and J.-B. Zuber, Commun. Math. Phys. {\bf 134}
(1990) 197.}%
\nref\dv{J. Distler and C. Vafa, Mod. Phys. Lett. {\bf A6} (1991) 259;
in {\sl Random Surfaces and Quantum Gravity}, Eds: O. Alvarez {\it et al}.}%
\nref\igor{D. Gross and I. Klebanov, {\it One-dimensional string theory on
a circle}, Nucl. Phys. {\bf B344} (1990) 475.}%

Matrix integrals of the above type have appeared in the work of the
mathematicians Harer and Zagier \harer\ and  Penner \penner\ in their
investigations of the Euler characteristic of the moduli space ${\cal M}_{g,s}$
of Riemann surfaces with $g$ handles and $s$ punctures. (See \iz\ for more
details on these wonderful calculations.) The double scaling limit of this
so-called Penner integral was considered by Distler and Vafa \dv\ who also
speculated on the relation with $c=1$ string theory. Their work has been
followed by a number of papers concerned with double scaling limits and
multi-critical behaviour of matrix models with logarithmic  potentials
\logkon. All these papers considered essentially the
case  $Z=1$ and $\tb_n=0$, in the notation of \sig.

Distler and Vafa noticed that --- after a double scaling limit and an analytic
continuation --- the Penner matrix integral could reproduce the $c=1$ partition
function at the self-dual radius $\beta=1$. Recall that the free energy at that
radius is given by \igor
\eqn\free{
{\partial^2 F\over \partial \mu^2} =
Re \int_0^\infty {dx \over x} e^{-i\mu x} \left({x/2 \over \sinh
x/2}\right)^2\ ,
}
and has an expansion
\eqn\expan{
 F = \half \mu^2 \log \mu - {1\over 12} \log \mu + \sum_{g=2}^\infty
(-1)^g {B_{2g} \over 2g(2g-2)} \mu^{2-2g}\ .
}
(Up to analytic terms in $\mu$.) This makes one wonder whether our result
\bloep\ can be sharpened to give the {\it unnormalized} correlation functions.

To this end let us put the incoming coupling constants $\tb_k$ to zero (and
thereby also $t_k=0$) and take a closer look at the integral
\eqn\integ{
\sigma(Z)= \int dY e^{i\mu Tr[Y Z^{-1} - \log Y]}\ .
}
 First of all it has a trivial $Z$-dependence
\eqn\zdep{
\sigma(Z) = (\det Z)^{N-i\mu} \cdot \sigma(1)\ .
}
Actually, it is convenient to work with the quantity $\FF$ defined by
\eqn\asymp{
e^\FF = (\pi i/\mu)^{-N^2 \over2} e^{-i\mu N} 2^{N/2}\cdot \sigma(1)\ .
}
As an asymptotic expansion in $1/\mu$
it has the representation
\eqn\rexp{
e^\FF = {\int dY \cdot e^{i\mu\sum_{k=2}^\infty {1\over k} Tr Y^k} \over
\int dY \cdot e^{i\mu {1\over 2} Tr Y^2}}\ .
}
This is known as the  Penner integral \penner\ and is usually considered
in `Euclidean signature', {\it i.e.} after analytic continuation $\mu=i \nu$,
$\nu$ real and positive.

The quantity $\FF$ has a beautiful geometrical
interpretation, calculating the virtual Euler characteristic of moduli space
in the open string field theory cell decomposition of moduli space. This is
essentially the same description of moduli space used by Kontsevich
\kontsevich. The expansion of $\FF$ reads
\eqn\blip{
\FF = \sum_{g=0}^\infty \sum_{s=1}^\infty N^s (-i\mu)^{2-2g -s} \FF_{g,s}\ .
}
(Here $s \geq 3$ in the case $g=0$.) The coefficients
are directly related to the Euler numbers
\eqn\euler{
\FF_{g,s} = \chi({\cal M}_{g,s})\ .
}

The $1/\mu$ asymptotics of the integral \rexp\ can be evaluated
using the methods described in \iz\ to give
\eqn\evalitz{
e^{\tilde F}=e^{-i \mu N} ({2 \pi i\over \mu})^{-N/2}
\bigl((-i\mu)^{i \mu}\Gamma(-i \mu)\bigr)^N\prod_{p=1}^{N-1}(1-p/i\mu)^{N-p}
}
from which one may obtain the formulae:
\eqn\fgs{
\FF_{g,s} = {(-1)^s B_{2g} \over 2g(2g-2+s)}{2g-2+s \choose s}\ ,
}
It is important that the terms with $s=0$, that is, the surfaces without
punctures, are absent.

We can explicitly do the summation over $s$ in \blip\ to obtain
\eqn\flip{
\FF = \sum_{g=0}^\infty \mu^{2-2g} \FF_g(N/i\mu)\ .
}
with
\eqn\glip{\eqalign{
\FF_g(x) & = {(-1)^g B_{2g} \over 2g(2g-2)} \left[1 - (1-x)^{2-2g}\right],
\qquad g \geq 2,\cr
\FF_1(x) & = -{1\over 12} \log (1-x),\cr
\FF_0(x) & = -\half (1-x)^2\log(1-x) +{3\over4} x^2 - \half x\ .\cr}
}
The double-scaling limit considered by Distler and Vafa in \dv\ keeps $N-i\mu$
fixed, while sending $N,\mu \rightarrow \infty$ (and $x \rightarrow 1$ in
\glip.).  This is clearly only possible for imaginary $\mu$, which is precisely
the  case they study. However, here we want to consider a simpler limit in
which  $\mu$ is kept fixed, but $N$ tends to infinity. We already mentioned
that  the parameterization \floep\ only makes sense in this limit. Indeed, the
absence of a double scaling limit is very much in the spirit of Kontsevich
integrals. The contribution for genus $2$ or higher have a smooth limit, as is
evident \ from \glip. (Recall, we send $x \rightarrow \infty.$)  However, we
have to worry about the genus zero and  one pieces, which have to be corrected
by hand. (This is by the way also true for the double scaling limit.)

Combining all ingredients we obtain the following final result for
the {\it unnormalized} generating functional for the $c=1$ string theory
\eqn\final{
\tau(t,\tb) = c(Z) \cdot \int dY \exp i\mu Tr\left[Y Z^{-1} -\log Y +
\sum \tb_k Y^{-k}\right]\ .
}
where the normalization constant is given by
\eqn\normal{\eqalign{
c(Z) = e^{-i\mu N} (2\pi i/\mu)^{N^2/2}&
(\det Z)^{i\mu-N} \cr
(1+iN/\mu)&^{{1\over2}(\mu + iN)^2 + {1\over 12}}
\mu^{{1\over2}\mu^2 - {1\over 12}} e^{{3\over4}N^2/\mu^2 - {i\over 2} N/\mu}
\ .\cr}
}
The expression \final\ has a smooth large $N$ limit.

\newsec{Other Backgrounds}

The results of the previous sections comprise in principle a calculation of the
partition function in arbitrary tachyon backgrounds (subject to the equations
of motion). The full space of classical backgrounds in the theory includes
in addition to these excitations of the ``discrete states'' corresponding
to global modes like the radius of the $1D$ universe and generalizations
thereof.
Of these, the ones best understood in terms of the matrix model are the
zero-momentum excitations which are thought to be
represented by variations in the double-scaled potential. In this section we
study the dependence of the amplitudes on these extra parameters.
We note that in principle the formulation of section three
applies in arbitrary potentials. What we add here is a study of the
variation of the reflection factor $R_q$, hence of the partition
function, under variations of the potential.

\subsec{Dependence on $\beta$}

The most obvious parameter is $\beta$, the radius at which we compactify the
scalar field $X$. The formulas of section four are valid for arbitrary
$\beta$, however as pointed out in
\kleblow, correlation functions at different radii are related. The
relation is most simply written in terms of rescaled couplings $t_n$. Defining
\eqn\redff{
\hat\CF[t_n,\tb_n;\beta;\mu] \equiv \mu^2 \CF[\mu^{{n\over 2\beta} -1} t_n,
\mu^{{n\over 2\beta} -1} \tb_n; \beta;\mu]
}
so that derivatives of $\hat\CF$ yield correlation functions of $\CT_q$, we
have
\eqn\kleblow{
\hat\CF[t_n,\tb_n;\beta;\mu] = {{1\over 2\beta}{\p\over\p\mu}\over
\sin\left({1\over 2\beta}{\p\over\p\mu}\right)}
\CF[\mu^{{n\over 2\beta} -1} t(n/\beta),
\mu^{{n\over 2\beta}-1}\tb(n/\beta);\infty;\mu]
\ .}

Comparing this with the $\beta\to\infty$ limit of the previous calculations is
a pretty consistency check.
As an example, set $\beta=1$ and consider the two-point function. Computing the
one-loop graph we find
\eqn\twopti{
{\p\hat\CF[t_n,\tb_n;1;\mu]\over\p t_n\p\tb_n} =
\mu^n \sum_{m=0}^{n-1} R^*_{p_m} R_{n-p_m}
= i^n \sum_{m=0}^{n-1} (-i\mu-m)_n
\ .}
Inverting the operator in \kleblow\ as
\eqn\invert{
{\sin({ \p\over 2\p \mu})\over {\p\over 2\p \mu}}
\hat \CF = \int_{-1/2}^{1/2} ds \hat \CF[\mu\to \mu+is]
}
we obtain
\eqn\twoptii{
\langle \CT_n \CT_{-n} \rangle_{\beta=\infty} =
i^n \int_0^n dx (\half-i\mu-x)_n
}
in agreement with the result of \mpr.

\subsec{Other zero-momentum modes}

The matrix model naturally suggests candidate representatives of the special
states at zero $X$ momentum. Operators with the appropriate quantum numbers
may be introduced as generating variations in the double-scaled potential
$V(\lambda)$. Their correlators may thus be studied by analysis of the
variation of the partition function $Z$ computed above under these changes in
$V$. From the definition of $I(q,\lambda_1,\lambda_2)$ we can obtain directly
constraints on the variation of $R_q$. Essentially these follow upon
integration by parts from the linear Gelfand-Dikii equation satisfied by a
product of Sturm-Liouville eigenfunctions
\ref\gd{I.M. Gel'fand and L.A. Dikii, {\it Asymptotic behaviour of the
resolvent
of Sturm-Liouville equations and the algebra of the Korteweg-de Vries
equations}, Russian Math. Surveys {\bf 30} (1975) 77.}. Explicitly, we have
\eqn\vircon{\eqalign{
L_{q,k} R_q &= 0 \qquad k\geq -1 \cr
L_{q,k} = -k(k^2-1){\p\over\p s_{k-2}} + 4iq&(k+1){\p\over\p s_k}
+2\sum_{p\geq 0} s_p (2k+p+2){\p\over\p s_{p+k}}\ ,\cr
}}
where the space of potentials is parametrized by the formal expansion
$V=\sum_{n\geq 0} s_n \lambda^n$. The operators $L_{q,k}$ for any fixed $q$ are
seen to satisfy the
commutation relations of (one half of) the Virasoro algebra. These were derived
from related considerations in
\ref\grdn{U. Danielsson and D.J. Gross, {\it On the correlation functions of
the
special operators in $c=1$ quantum gravity}, Nucl. Phys. {\bf B366} (1991) 3.};
the details of the derivation in the
present context as well as the relation to this work appear in the appendix.

Via \genfun\ these imply constraints on the $V(\lambda)$-dependence of the
partition function, since this arises only through $R_q$.
These however are nontrivial to write down explicitly. In particular,
we note that they do not seem to fit into the $W_{1+\infty}$ algebra
discussed in section four. Furthermore, it is easy to see that away from
$s_k=0$ the
identity \useful\ ceases to hold. Thus perturbing away from the standard
background may break the symmetry of section four. Further work is
required to clarify the relation of the
various symmetry algebras which appear in this model.

\newsec{Discussion}

Some remarks on the $W$-constraints \wconstr\ are in order. First, it cannot
have escaped the reader that these
constraints are strongly reminiscent of the
famous $W$-constraints of $c<1$ models coupled to
gravity
\refs{\dvv,\kawai}.
In these latter models there is only one continuous
spacetime coordinate and there is only one set of
couplings $t_j$ rather than the $t,\bar{t}$ of the
$c=1$ model. Moreover, closer comparison of the identities
reveals some important differences. For example, at
$c<1$ the $t_n$ couple to an infinite set of gravitational
descendents, while at $c=1$ the $t,\tb$ couple to
gravitational primaries. Nevertheless, a clearer
spacetime interpretation of the $c<1$ models will
probably emerge from a comparison of these identities.%
\foot{In
\ref\awada{M.A. Awada and S.J. Sin, {\it Twisted $W_\infty$ symmetry
of the KP hierarchy and the string equation of $d=1$ matrix models},
Univ. Florida preprint IFT-HEP-90-33; {\it The string difference
equation of $d=1$ matrix models and $W_{1+\infty}$ symmetry
of the KP hierarchy}, IFT-HEP-91-3.} proposals for $c=1$ flow
equations were made by taking the $N\to \infty$ limit of the
$W_N$ constraints of the $c<1$ models. It should be noted that,
although our equations have some similarities to the proposals
of \awada, they are not equivalent.}

{}From the relation of these results to a
Kontsevich-type matrix model it appears that we have
taken a step closer to a unified description of all
the $c\leq 1$ models along the lines
proposed by \refs{\morozov,\robbert}. Moreover,
the description \final\ of the partition function
is a strong hint that the c=1 correlators have a
description in terms of a topological field theory.
If this is so then the present results provide a
direct bridge between a topological field theory
at the self-dual radius and the local physics of
the $c=1$ tachyon in the uncompactified theory.

There have been many discussions of $W_\infty$ symmetry
in the $c=1$ system. Our constraints are related to
the results of
\nref\wvii{D. Minic, J. Polchinski, and Z. Yang, {\it Translation-invariant
backgrounds in 1+1 dimensional string theory},
Nucl. Phys. {\bf B362} (1991) 125.}%
\nref\wviii{J. Avan and A. Jevicki,
{\it Classical integrability and higher symmetries of collective field
theory}, Phys. Lett. {\bf 266B} (1991) 35;
{\it Quantum integrability and exact eigenstates of the collective string
field theory}, Phys. Lett. {\bf 272B} (1991) 17.}%
\nref\wix{S.R. Das, A. Dhar, G. Mandal, S. R. Wadia,
{\it Gauge theory formulation of the c=1 matrix model:
symmetries and discrete states}, IAS preprint IASSNS-HEP-91/52, hepth 9110021;
{\it Bosonization of nonrelativistic fermions and $W_\infty$ algebra},
Mod. Phys. Lett. {\bf A7} (1992) 71\semi
A. Dhar, G. Mandal, S. R. Wadia,
{\it Classical Fermi fluid and geometric action for $c=1$}, Preprint
IASSNS-HEP-91-89, hepth 9204028;
{\it Non-relativistic fermions, coadjoint orbits of $w_\infty$ and string field
theory at $c=1$}, Preprint  TIFR-TH-92-40, hepth 9207011.}%
\refs{\msi,\wvii{--}\wix}.
The other modes of
the $W_\infty$ currents appearing in equation
\wconstr\ define a set of operators
$\sigma_n(T_q)$ whose correlation functions are
determined by the subleading terms proportional to
$\lambda^{-|q|-2n}$ in the large $\lambda$ asymptotics of
the eigenvalue correlators.%
\foot{In \moore\ these operators
were denoted $\sigma_{2n}(\CO_q)$. In
\msi\ it was pointed out that they only have
contact term interactions.}
These ``operators'' exist at any radius for $X$ and have
free fermion representations as fermion bilinears.
Their correlators are
also given by a Toda tau function generalizing that in
\genfun. Note that these operators
appear at any momentum $q$ and are related to fractional
powers of $\ell$ (or, equivalently, of $\lambda$).
Therefore, at generic $q$ they {\it cannot} be the special
state operators but rather are related to contact terms
associated to singular geometries created by intersecting
macroscopic loops \msi. At integer $q$ the distinction
between special states and the $\sigma_n(T_q)$ is less
clear. We hope to return to the subtleties of these
contact terms in a future publication.

The $W_\infty$ symmetry we have discussed might also be related to
the $W_\infty$ Ward identities of
\nref\wx{E. Witten, {\it Ground ring of two dimensional string theory},
Nucl. Phys. {\bf B373} (1992) 187.}%
\nref\polyakov{I.R. Klebanov and A.M. Polyakov, {\it Interaction of
discrete states in two-dimensional string theory}, Mod. Phys. Lett. {\bf A6}
(1991) 3273.}%
\nref\kms{D. Kutasov, E. Martinec, and N.
Seiberg, {\it Ground rings and their modules
in 2d gravity with $c\le 1$ matter}, Phys. Lett. {\bf 276B} (1992) 437.}%
\nref\klebbb{I. Klebanov, {\it Ward identities in two-dimensional
string theory}, Mod. Phys. Lett. {\bf A7} (1992) 723.}%
\nref\tanii{Yoichiro Matsumura, Norisuke Sakai, Yoshiaki Tanii,
{\it Interaction of tachyons and discrete states in $c = 1$ 2-d quantum
gravity}, hepth 9201066}%
\nref\witzwie{E. Witten and B. Zwiebach, {\it Algebraic structures
and differential geometry in 2d string theory}, Nucl. Phys.
{\bf B377} (1992) 55.}%
\nref\erik{E. Verlinde, {\it The master equation of 2D string theory},
IAS preprint IASSNS-HEP-92/5, hepth 9202021.}%
\refs{\wx{--}\erik}. In these references the Liouville field
is treated as a free field, in other words, one works at $\mu=0$.
One should be cautious about
identifying these $W_\infty$ symmetries with those of
the matrix model. As we have emphasized, the
$W_\infty$ modes of the matrix model $\sigma_n(T_q)$ are constructed from
the tachyon degrees of freedom in distinction to the $W_\infty$
currents of \refs{\wx{--}\erik}. Moreover, our Ward identities are
highly nonlinear when expressed in terms of the correlation functions%
\foot{This is already true at genus zero \mrpl.} in
contrast to the quadratic identities of \refs{\kms{--}\erik}. Finally the
ghost sector of the theory is crucial in
\refs{\kms{--}\erik}, leading to many more ``special state
operators'' at given $X,\phi$ momenta than are considered in
\refs{\msi,\wvii{--}\wix}. Clearly there is a certain
amount of tension between these two approaches and
further work is needed to see if these differences are
superficial or essential.

We must emphasize that at $c=1$ the $W_\infty$-constraints are actually
somewhat secondary, since we have an explicit solution of
the appropriate Toda tau function given by
\genfun. Analogous representations for the $c<1$ tau functions
(at nontopological points) replace the simple operator
$S$ by complicated and uncomputable objects like the ``star operators'' of
\ref\mrcrg{G. Moore, {\it Geometry of the string equation},
Comm. Math. Phys. {\bf 133} (1990) 261;
{\it Matrix models of 2d gravity and isomonodromic deformation},
in: {\sl Common Trends in Mathematics and Quantum Field Theory}
Proc. of the 1990 Yukawa International Seminar, Kyoto.  Edited by T.
Eguchi, T. Inami and T. Miwa.  To appear in the Proc. of the
Cargese meeting, Random Surfaces, Quantum Gravity and Strings, 1990.}.
This is why the Virasoro constraints at $c<1$ are essential to the
actual computation of amplitudes.

It would be interesting to investigate further the
physical properties of these different {\it time-dependent} backgrounds.
In
\refs{\mrpl,\sg}
some results along these lines were discussed.
Our result \genfun\ should
allow a much more complete analysis of the space of
time-dependent backgrounds in 2D string theory and
the various phase transitions occurring as one increases
the coordinates $t_k$. What is needed for further
progress is a more effective way to compute the
tau function (perhaps from the Kontsevich representation)
or a deeper understanding of the infinite dimensional
geometry of the associated Grassmannian.

\bigskip
\centerline{\bf Acknowledgements}

We would like to thank T. Banks, E. Martinec, N. Seiberg, C. Vafa,
and H. Verlinde for discussions.
This work is supported by DOE grant DE-AC02-76ER03075 and
by a Presidential Young Investigator Award (G.M., R.P.)
and by the W.M. Keck foundation (R.D.). G.M would like to thank the
Isaac Newton Institute for Mathematical Sciences for
hospitality.

\appendix{A}{Dependence on the potential}

In this appendix we will derive constraints on the dependence of the free
energy upon the double-scaled matrix model potential $V(\lambda)$. We restrict
attention to variations of the potential which preserve the asymptotics
$V(\lambda) \sim -{1\over 4}\lambda^2$ for large $\lambda$. For convenience,
we will assume that the asymptotic behavior of $V$ at $\lambda\to -\infty$ is
such that the eigenfunctions decay rapidly enough to render all integrals
convergent and all boundary terms negligible. To all orders of perturbation
theory, of course, these details are irrelevant.

We wish to compute the change in the correlators \nptev\ under a variation
$V(\lambda)\to V + \delta V(\lambda)$ preserving the asymptotics. The change in
\nptev\ is clearly computed from the variation of the resolvent $I$. Thus write
\eqn\vari{
\delta I(q,\lambda_1,\lambda_2) =
\langle\lambda_1 |\delta\left( {1\over H-\mu-iq} \right) |\lambda_2\rangle =
-\int_{-\infty}^\infty d\lambda \delta V(\lambda) I(q,\lambda_1,\lambda)I
(q,\lambda,
\lambda_2)
\ .}
We now recall the calculation of $I$ from \mpr\ (see appendix A of this work
for a detailed calculation for particular potentials).
For simplicity let $q>0$.
We will make use of the eigenfunctions of $H={d^2\over d\lambda^2}-V(\lambda)$
with eigenvalue $z=\mu+iq$.
We have for any $V$ with the correct asymptotics two solutions
\eqn\chis{
\chi^\pm(z,\lambda)
{\buildrel \lambda\to +\infty\over\sim} \lambda^{-\half \mp iz}
e^{\pm i\lambda^2}
\ .}
In terms of these we can write the resolvent quite easily by imposing the
boundary conditions and the defining property $(H-z)I(q,\lambda_1,\lambda_2) =
\delta (\lambda_1-\lambda_2)$ as in \mpr
\eqn\i{
I(q,\lambda_1,\lambda_2) = -i\theta (\lambda_1-\lambda_2) \biggl[
\chi^-(z,\lambda_1)\chi^+(z,\lambda_2) +
R_q \chi^-(z,\lambda_1)\chi^-(z,\lambda_2)\biggr] + (\lambda_1\leftrightarrow
\lambda_2)
\ .}
The reflection factor $R_q$ contains all the effects of the potential, and for
the standard $V$ is given by \rq.
Inserting this into \vari\ and neglecting terms of order $\delta
V(\lambda_{1,2})$ for large $\lambda_i$, we find that a variation of $V$
yields
\eqn\varr{
\delta R_q = -i \int_{-\infty}^\infty d\lambda \delta V(\lambda)
\psi(z,\lambda)^2
}
where $\psi = \chi^+ + R_q\chi^-$ is the solution satisfying the boundary
conditions at small $\lambda$.
The integrand $F(z,\lambda) = \psi(z,\lambda)^2$
in \varr\ satisfies a differential equation
\gd\ following  from that satisfied by $\psi$
\eqn\egd{
 F''' - 4(V(\lambda)+z)F' -2V' F = 0
}
where primes denote $\lambda$ differentiation. Let us choose as a convenient
set of variations of the potential $\delta V(\lambda) = \epsilon e^{- \ell
\lambda}$. Inserting this in \varr\ and integrating by parts we find%
\foot{The similarity of this to the WdW equation of \moore\ is no coincidence;
setting $z=\mu$ and $\lambda_1=\lambda_2$ we find that \vari\ is essentially
the WdW wavefunction of the cosmological constant.}
\eqn\ew{
[ \ell^3 - 4 z \ell - 4 \ell V(-{d\over d\ell}) + 2 V'(-{d\over d\ell}) ]\delta
R_q = 0
\ .}
The integration by parts is justified by the limiting conditions we have
imposed upon $\psi$ and $\delta V$.

 Formally expanding $V=\sum_{n\geq 0} s_n \lambda^n$ the bounce factor
becomes a function of the $s_j$: $R_q=R_q[s_1,s_2,\dots]$. Rewriting $\delta V$
as a motion in $s_j$ and inserting the resulting expression for $\delta R_q$ in
\ew\ we obtain (after shifting $s_0$)
\eqn\vir{\eqalign{
L_{q,k} R_q &= 0 \qquad k\geq -1 \cr
L_{q,k} = -k(k^2-1){\p\over\p s_{k-2}} + 4iq&(k+1){\p\over\p s_k}
+2\sum_{p\geq 0} s_p (2k+p+2){\p\over\p s_{p+k}}\ .\cr
}}
These constraints were obtained in \grdn\ by different means. We will show that
the two results are equivalent, but note here that the present derivation has
the advantage of working with potentials with the correct asymptotics
throughout, as well as demonstrating explicitly the justification for the
various integrations by parts.

The Virasoro constraints in \grdn\ were obtained as differential equations for
the cosmological constant one-point function. We have obtained above
identical constraints on $R_q$, from which one can derive constraints on the
partition function. We will now relate the two quantities, demonstrating that
in fact the two sets of constraints coincide.
Begin with the formula for the two-point function \mpr:
\eqn\twpti{\langle T_q T_{-q}\rangle =\int_0^q dx R_{x} R_{q-x}^*
}
or its differentiated version:
\eqn\twptii{{\p\over \p \mu}\langle T_q T_{-q}\rangle=
2Im\bigl[R_qR_0^*\bigr]
}
 For $q>0$ we may write:
\eqn\ebf{
R_q=e^{i\Theta(\mu+i q;V)}
}
where for $E$ real the pure phase $e^{i\Theta(E;V)}$ is the reflection
coefficient for a free fermion of energy $E$ in the double-scaled matrix
potential $V(\lambda)$.

We now obtain the specific heat from the limit as $q\to 0$.
As explained in
\nref\joei{J. Polchinski, {\it Critical Behavior of random surfaces in
one dimension}, Nucl. Phys. {\bf B346} (1990) 253.}%
\nref\nati{N. Seiberg, {\it Notes on quantum liouville theory and
quantum gravity}, in {\sl Common Trends in
Mathematics and Quantum Field Theories}, Proceedings of the 1990 Yukawa
International Seminar, Prog. Theor. Phys. Supp. {\bf 102}.}%
\nref\kdf{D. Kutasov and Ph. DiFrancesco, Phys. Lett. {\bf 261B}(1991)385;
D. Kutasov and Ph. DiFrancesco, {\it World Sheet and Space Time
Physics in Two Dimensional (Super) String Theory}, Princeton preprint
PUPT-1276, hepth 9109005.}%
\refs{\joei,\nati,\kdf}
the cosmological constant vertex operator in the $c=1$ theory is
given by
\eqn\ccop{
T_0=\int_\Sigma \phi e^{\sqrt{2}\phi}
}
and is therefore obtained by the limit
\eqn\lmv{
\lim_{q\to 0}{1\over q}T_q=T_0 \quad .}

Taking the limit in \twptii\ we find the leading
order begins at $q^2$, as expected from \lmv. Indeed, quite
generally, the low energy theorem of \mpr\ shows that for
$n>2$-point functions if $k<n$ momenta $q_i$ approach
zero the amplitude behaves like
$\prod_{i=1}^k q_i\bigl({\p\over \p \mu}\bigr)^k\langle \prod_{k+1}^n
V_{q_i}\rangle$ in accordance with the expectations of Liouville theory.
Thus, the apparent $ {1\over q}$ divergence in \lmv\ does not
appear. From the Liouville point of view this may be interpreted as
the decoupling of the
wrong branch dressing of the vertex operator  \refs{\nati,\kdf}.
The low energy theorem is more subtle in the case $n=2$.
In this case the leading order behavior is $q+\CO(q^2)$.
The first term is $\mu$-independent and physically sensible,
being the inverse on-shell propagator at genus zero. The second
term of order
$\CO(q^2)$ defines the correct zero-momentum two-point function.

Taking the limit of \twpti\ and bearing in mind the
above remarks we obtain the equation
\eqn\twocc{
\langle T_0 T_0 \rangle = -\Theta'(\mu;V)
}
and hence the nonperturbative one-point function and vacuum energy are
\eqn\vac{
\langle T_0\rangle=-\Theta(\mu;V)=i\log R(\mu;V)
\ .}

\listrefs
\listfigs
\bye